%% file: paper.tex
\documentclass[10pt,twocolumn]{article}
\usepackage[top=1.9cm, bottom=1.9cm, left=1.8cm, right=1.8cm]{geometry}
\usepackage{times}  %
\usepackage[sort&compress,numbers]{natbib}  %
\usepackage[hyphens]{url}
\usepackage{graphicx}  %
\usepackage[keeplastbox]{flushend}
\usepackage{graphicx} %
\usepackage{tikzsymbols}
\let\oldbibliography\thebibliography
\renewcommand{\thebibliography}[1]{%
\oldbibliography{#1}%
\setlength{\itemsep}{2pt}%
}
\usepackage[hang,flushmargin]{footmisc}

\usepackage[compact]{titlesec}
\titlespacing*{\section}{0pt}{*3}{3pt}
\titlespacing*{\subsection}{0pt}{*2}{2pt}
\usepackage{xspace}
\makeatletter
\def\url@leostyle{%
  \@ifundefined{selectfont}{\def\UrlFont{}}%
  {\def\UrlFont{}}%
}
\makeatother
\usepackage[hyphenbreaks]{breakurl}
\usepackage[bookmarks=true, bookmarksnumbered=true, colorlinks=true, linkcolor=linkcol, citecolor=citecol, urlcolor=urlcol, hypertexnames=true]{hyperref}
\definecolor{darkgreen}{RGB}{0, 100, 0}
\definecolor{linkcol}{rgb}{0.3,0,0}
\definecolor{citecol}{rgb}{0.3,0,0}
\definecolor{urlcol}{rgb}{0.3,0,0}
\definecolor{vlightgray}{gray}{0.925}
\usepackage[small,bf]{caption}
\usepackage{tikz}
\usepackage{comment}
\usepackage{booktabs}
\usepackage{wrapfig}
\usepackage{xspace}
\usepackage{mdframed}
\usepackage{fdsymbol}
\usepackage{url}
\usepackage{algorithmic}
\usepackage{graphicx}
\usepackage{textcomp}
\usepackage{tablefootnote}
\usepackage{threeparttable}  
\usepackage{multirow}
\usepackage{bm}
\usepackage{bmpsize}
\usepackage{xcolor}
\usepackage{lipsum}
\usepackage{lscape}
\usepackage{tabularx}
\usepackage{multicol}
\usepackage{multirow}
\usepackage{adjustbox}
\usepackage{balance}

\makeatletter
\def\url@leostyle{%
  \@ifundefined{selectfont}{\def\UrlFont{\small}}%
  {\def\UrlFont{}}%
}
\makeatother
\usepackage{etoolbox}

\newcommand{\refappendix}[1]{\hyperref[#1]{Appendix~\ref*{#1}}}

\makeatletter
\newcommand{\linebreakand}{%
  \end{@IEEEauthorhalign}
  \hfill\mbox{}\par
  \mbox{}\hfill\begin{@IEEEauthorhalign}
}
\makeatother

\begin{document}
\date{}

\title{Difficult for Thee, But Not for Me: Measuring the Difficulty and User Experience of Remediating Persistent IoT Malware}

\author{{\rm Elsa Rodr\'{i}guez}, {\rm Max Fukkink, Simon Parkin, Michel van Eeten, Carlos Ga\~{n}\'{a}n}\\
Delft University of Technology\\
\textit{\{e.r.turciosrodriguez\},}\textit{ m.c.s.fukkink@student.tudelft.nl,}\\\textit{\{s.e.parkin, m.j.g.vaneeten, c.hernandezganan \}@tudelft.nl}}

\maketitle
\input{abstract}

\pagestyle{plain}

	\input{introduction}

	\input{background}

	\input{methodology}

	\input{ethics}

	\input{results}

	\input{related_work}
	\input{discussion}
	\input{conclusion}

	\input{acknowledgements}
	
%|||||||||||||||||||||||||||||||||||||||||||||||||||||||||||||||||||||||||%
%               References
%|||||||||||||||||||||||||||||||||||||||||||||||||||||||||||||||||||||||||%
\balance

\bibliographystyle{abbrv}

%\bibliography{references}

\input{paper.bbl}
%|||||||||||||||||||||||||||||||||||||||||||||||||||||||||||||||||||||||||%
%               Appendices (if any)
%|||||||||||||||||||||||||||||||||||||||||||||||||||||||||||||||||||||||||%

\input{appendix}

\end{document}

%% file: abstract.tex
% !TeX root = paper.tex
\begin{abstract}

Consumer IoT devices may suffer malware attacks, and be recruited into botnets or worse.
There is evidence that generic advice to device owners to address IoT malware can be successful, but this does not account for emerging forms of persistent IoT malware. Less is known about persistent malware, which resides on persistent storage, requiring targeted manual effort to remove it. This paper presents a field study on the removal of persistent IoT malware by consumers. We partnered with  an  ISP  to  contrast remediation  times  of  760 customers across three malware categories: Windows malware, non-persistent IoT malware, and persistent IoT  malware.  We  also contacted ISP customers identified as having persistent IoT malware on their network-attached storage devices, specifically QSnatch. We found that persistent IoT malware exhibits a mean infection duration many times higher than Windows or Mirai malware; QSnatch has a survival probability of 30\%  after 180 days, whereby most if not all other observed malware types have been removed.
For interviewed device users, QSnatch infections lasted longer, so are apparently more difficult to get rid of, yet participants did not report experiencing difficulty in following notification instructions. We see two factors driving this paradoxical finding: First,  most  users  reported  having  high  technical competency. Also, we  found  evidence  of  planning behavior  for  these  tasks and the need for multiple notifications. 
Our findings demonstrate the critical nature of interventions from outside for persistent malware, since automatic scan of an  AV tool or a power cycle, like we are used to for Windows malware and Mirai infections, will not solve persistent IoT malware infections.

\end{abstract}

%% file: introduction.tex
% !TeX root = paper.tex
\section{Introduction}\label{sec:introduction}

Smart home devices keep proliferating and, unfortunately, so do the malware families targeting these devices. Solutions for malware detection and removal have a long lineage, going back at least two decades. After the chaos of the global virus outbreaks of the early 2000s, countermeasures slowly started to emerge from what became the anti-virus industry and from operating system manufacturers like Microsoft. Years of painstaking development have resulted into the highly automated and usable tools that consumers rely on today to detect and remediate infections on their personal computers. 

Then, about five years ago, IoT malware surged, most notably in the form of the Mirai family~\cite{Antonakakis2017}. It captured millions of surveillance cameras, digital video recorders, routers, and many more devices that researchers could not identify \cite{rodriguezsuperspreaders}. Here, none of our automated tools work and many of the hard-earned usability lessons for PC malware cannot be applied. These devices are typically headless, lack a graphical user interface (GUI) or a peripheral device for users to learn about an infection and take the recommended actions. There are no standard anti-virus tools that can run on these devices. (As an aside, some vendors are now offering dedicated anti-IoT malware devices which users are meant to put in their local network. Bundled with a subscription, they can cost hundreds of dollars per year, which explains why they currently are niche products. These devices can potentially do detection based on network traffic, but not remediation of the infection. The latter task remains with the user.) 
To make matters worse, IoT represents an enormously heterogeneous population of devices in terms of design and function \cite{Kumar2019}. The deployment of tens of thousands of different devices makes it all but impossible to give users security advice that is actionable for their specific devices. 

Industry and governments have been coping with this challenge by providing consumers with highly generic instructions that try to cover all manner of devices and attack vectors \cite{NationalCybersecurityCentre2019,NationalInitiativeforCybersecurityCareers2019}. This advice suffers from usability problems, since it could not be made actionable for specific device, leaving consumers to figure out how to take actions like disabling Telnet, changing a factory-default password or installing a firmware update. Surprisingly, these coping strategies did have some success. 

Remediation levels were found to be high \cite{Cetin2019}, even though the security advice was poorly understood by users and it lacked a deterministic path to removing the infections \cite{bouwmeester2021thing}. This success was helped by fleeting nature of the infections. The bulk of all IoT malware resides in memory only and does not gain a persistent foothold on the device. Thus, a power cycle would remove it---albeit only temporarily if not combined with other protection measures like a password change.

Now the next challenge has arrived: persistent IoT malware~\cite{bock2019assessing,brierley2020persistence,Fbot,Matryosh}. It combines the worst of both worlds: the persistence of PC malware with absence of effective and usable detection and removal tools of IoT malware. Does persistent IoT malware make remediation more difficult? How do users experience their remediation efforts? Learning the answers to these questions is critical in responding to the next evolution of IoT malware.

This paper presents the first field study on removing persistent IoT malware by consumers. We partnered with an Internet Service Provider (ISP) in The Netherlands to compare the remediation times of 760 customers for three categories of malware families: persistent Windows malware, non-persistent IoT malware and persistent IoT malware. In the latter family, we focus on QSnatch, also known as Derek~\cite{NationalCybersecurityCentre2019}, as a case study. We selected QSnatch since it was the most prevalent IoT malware family, which was not memory resident only, in the network of our  partner ISP at the time of this study. Besides, QSnatch is an appropriate representation of a persistent IoT threat for several reasons. First, according to the US Cybersecurity Infrastructure Security Agency (CISA) and the National Cyber Security Centre UK (NCSC-UK), the number of QSnatch reported infections grew from 7,000 devices in October 2019 to more than 62,000 in June 2020 \cite{CybersecurityandInfrastructureSecurityAgency,cimpanu2020}. Second, non-profit organization Shadowserver %has reported similar numbers \cite{TECHMONITOR} 
recently reported QSnatch as the second most important threat after Mirai---in some countries even as the top IoT malware family \cite{Shadowserver}. Finally, as highlighted by \cite{272224} network attached storage (NAS) are among the top devices targeted by IoT malware. 

Next, we contacted ISP customers who had suffered from a QSnatch infection in the past year.
We interviewed 57 customers with an instrument design informed by the COM-B behavior model \cite{michie2011}, which stresses the importance of individuals' capabilities, motivations, and opportunities to perform a behavior. The model has been suggested to be applied to understand behavior change in security \cite{enisa2018cybersecurity}. We also compared the cleanup success of interviewees to the non-interviewed QSnatch victims. 

Overall, we find that, yes, persistent IoT malware is more difficult to remediate. These infections last more than three times longer
than infections with Windows malware or non-persistent IoT malware, namely Mirai. 
This is consistent with the fact that the remediation for QSnatch consists of a convoluted series of steps.
Surprisingly, though, the interviewed users reported that they did not find the remediation steps particularly difficult.

We see two factors driving this paradoxical finding. First, most users reported having high technical competency---in fact, the majority even reported working as an IT professional. So their tolerance for difficult tasks is a lot higher than for the average user. Their frame of reference might be complex IT admin tasks, rather than the more simple consumer action of running an AV tool. We found evidence of planning behavior for these tasks, e.g., doing it on the weekend. There might be a self-selection process at work, owners of network-attached storage (NAS), as a new technology, are much more likely to be technically competent \cite{dedehayir2020innovators}, thus experiencing the difficult task as a normal task, but then they do need some time and effort to execute it. 

The second factor that explains why users did not find it very difficult, yet they took longer to remediate than Windows and Mirai infections, is that the latter can also get remediated without user action. An automatic scan of an AV tool or Windows malware removal might find and remove the infection, without the user even noticing. For Mirai infections, a power cycle removes the infection (even though it leaves open the possibility of reinfection). Such `natural' remediation is not possible for QSnatch. Only user action can get rid of it.

In sum, we make the following main contributions: 

\begin{itemize}
    \item We quantify the infection duration for 228 customers infected with persistent IoT malware, and compared it to customers infected with memory-resident IoT malware and Windows malware.  Compared to Windows malware, the mean infection time of persistent IoT malware is three times higher. Compared to  memory-resident IoT malware, persistent IoT malware mean infection time is five times higher.
    \item We estimate the survival probability of different types of malware and statistically compute differences between malware families. Our results show that 30\% of the infected subscribers with persistent malware remain infected even after six months since the infection was detected.
    \item We provide real-world evidence of users mitigating persistent IoT malware. Our results show that all QSnatch-infected customers remediate right or closely after receiving a notification. 
    \item We derive a set of recommendations to expedite the cleanup processes of persistent IoT malware based on the interviews conducted with 57 infected customers. 

\end{itemize}

\vspace{-5pt}

%% file: background.tex
% !TeX root = paper.tex
\section{Background}\label{sec:background}

\subsection{QSnatch and persistent malware}

Most popular IoT malware families, such as Mirai in its many variants, are stored within the temporary file systems of the IoT devices. They resided in the Random-Access Memory (RAM) of devices. This memory is volatile, thus any malware residing in it will be removed from the device if the device is powered off or just restarted. Persistent malware, on the other hand, is stored among system files of the operating system, they can be added to the startup process of the operating system or schedule processes, being able to survive reboots, and maintaining a connection with the device to keep it as part of a network of bots.

Our study focuses on an important example of persistent IoT malware called QSnatch. QSnatch targets network-attached  storage (NAS) devices from the manufacturer QNAP \cite{QNAP}. Some characteristics that make QSnatch persistent are that it changes scheduled tasks of the device, prevents firmware updates by rewriting the URL from where the update comes from, and steals usernames and passwords \cite{Davis}. 
The malware uses Domain Generation Algorithms (DGA) to communicate with the command and control servers controlled by the attackers \cite{Casino2021}. 

Different security firms have characterized the capabilities of QSnatch \cite{SoftwareTested, cimpanu2020}: 

\begin{itemize}
    \item Common gateway interface (CGI)  password logger - a fraudulent version of the device admin login page, recording  authentications and passing them to the legitimate login page.
    \item Credential scraper.
    \item SSH backdoor - Allowing to execute arbitrary code on a device.
    \item Exfiltration - When run, QSnatch steals a predetermined list of files, which includes system configurations and log files. These are encrypted with the attacker's public key and sent to their infrastructure over HTTPS.
    \item Webshell functionality for remote access. 
\end{itemize}

QSnatch poses a threat to users besides the possibility of being used for Distributed Denial of Service (DDoS) attacks or to deliver malware payloads to other devices.

\subsection{QSnatch remediation mechanisms} \label{sec:remediation_mechanism_manuf_ISP}

To remediate QSnatch, QNAP has published a series of recommended steps. Our partner Internet Service Provider (ISP), in turn, created a shorter version of these steps to include in their notifications to affected customers. 

\subsubsection {Manufacturer recommendation}

QNAP's security advisory to address QSnatch infections recommends a whopping 84 user actions, organized around several high-level steps~\cite{QNAP2020}:
\begin{itemize}
    \item Update QNAP turbo station (QTS) 
    to the latest available version.
    \item Install and update Malware Remover to the latest version.
    \item Install and update Security Counselor to the latest version.
    \item Update your installed QTS applications to the latest versions if available in the App Center.
    \item Configure settings to enhance system security.
\end{itemize}

Each of these steps includes actions like changing various settings of the device, enabling and disabling features, changing passwords and configurations, and subscribing to QNAP Security Newsletters \cite{QNAP2020}. 

Different than how Mirai could, in practice, be removed by resetting an infected device \cite{Cetin2019, bouwmeester2021thing, rodriguez2021user}, resetting a NAS would not lead to remediation. In contrast to Windows malware, where users may count on existing tools to remove infections in the background, such as antivirus software, removing QSnatch requires recognizing the correct information and applying the security advice. To solve the issue, users need to perform more steps than for removing Mirai malware \cite{Cetin2019, bouwmeester2021thing, rodriguez2021user} or running antivirus, and if these tasks are perceived as challenging or dull, users might postpone them \cite{van2000procrastination}; thus, making this infection more difficult to remediate.

\subsubsection{Internet Service Provider recommendation} 

The partner ISP contacts customers who suffer from a QSnatch infection. The notification includes the recommended steps for remediation. Rather than point customers to the complicated advisory on the QNAP website, the ISP has condensed the advisory into a shorter and simplified version of the remediation process.

The notification explains to the user that a QNAP network-attached storage device has been compromised with QSnatch malware and then provides nine steps to solve the infection (see \refappendix{sec:appendixA} for the full notification). Since QSnatch has the capability of rewriting the URL for downloading the new firmware and blocking the launch of the QNAP Malware Remover tool\cite{Davis}, the ISP recommends to users to do the following:

\begin{itemize}
    \item Go to the website: \url{qnap.com/en-en/download}
    \item Under ``1 - Product type", select the option ``NAS / Expansion" and select the number of slots present on the right.
    \item Under ``3-Model", select the type of NAS you are using.
    \item Under the ``Operating System" tab, select the most recent version and download it via the ``[REGION]" button.
    \item Open the NAS on your PC or Mac and choose firmware update, and then Manual update.
    \item Browse to the downloaded file and update the firmware / operating system.
    \item Go to APP Center and choose ``Malware Remover" and download it on your PC or Mac.
    \item Click on ``manual update" in App center, browse to the download file and update the Malware Remover.
    \item Run a scan with the Malware remover.
\end{itemize}

\subsubsection{Notification process} 

At our partner Internet Service Provider (ISP), the starting point to handle all infections is a feed from a third party, Shadowserver, specifically the Drone Report \cite{Shadowserver2021}. Shadowserver is a non-profit security organization that shares abuse data to make the Internet secure. It is a trusted source for network providers, national governments and law enforcement \cite{shadow}. 
The Drone Report is received daily by the ISP abuse handling department and contains data on infections for many different malware families. 

It includes the Internet Protocol (IP) addresses where infected machines were observed. These addresses were captured by different techniques such as sinkholes, darknets, honeypots and other sources \cite{Shadowserver2021}. The ISP connects the Shadowserver IP addresses with their customers' data. Once the affected customers are identified, an automated system sends an email notification about the detected security issue. These notifications can be customized to the type of malware. So for Windows malware, users get different instructions than for Mirai IoT malware.

Shortly after  QSnatch was added to the Shadowserver Drone Report, the ISP included these infections in the standard abuse handling workflow for all infections. Affected users would receive the email with the ISP's customized recommendations for removing QSnatch.

%% file: methodology.tex
% !TeX root = paper.tex
\section{Methodology}\label{sec:methodology}

\begin{figure*}[t!]
	\vspace{-10pt}
	\centering
	\includegraphics[width=1.65\columnwidth]{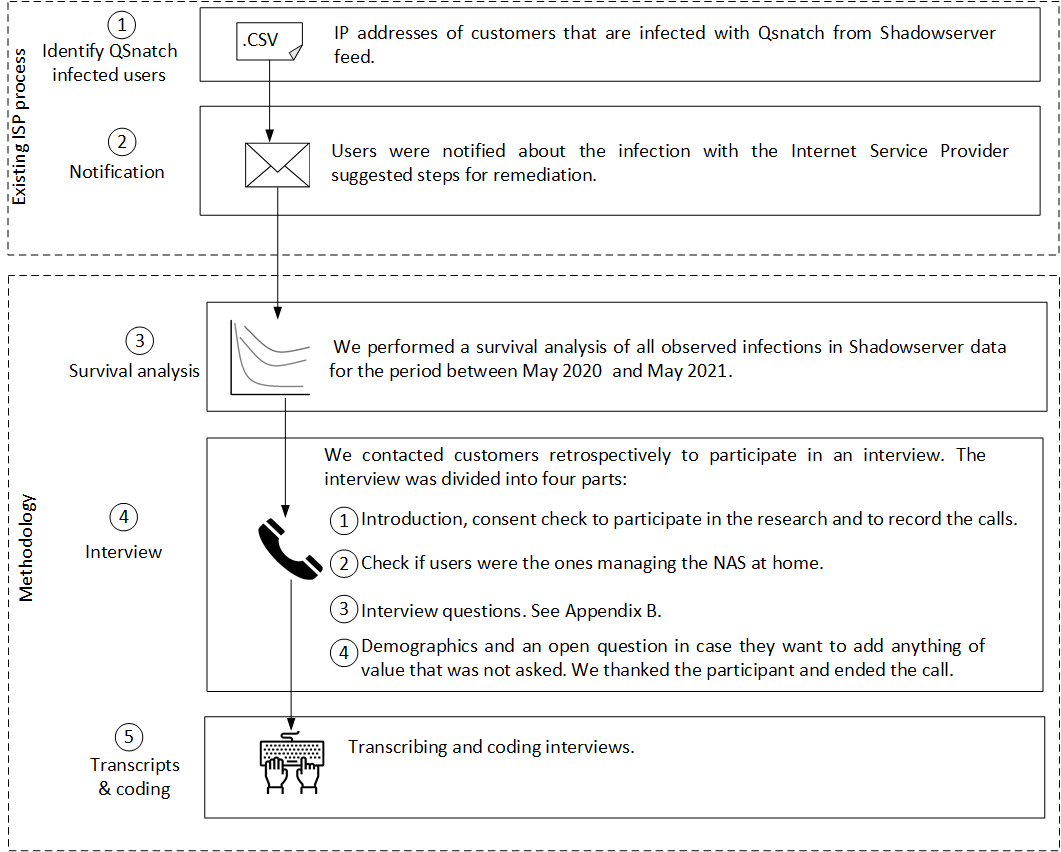}
	\caption{The ISP existing process and overview of the mixed-method approach} 
	\label{fig:Method}
	\vspace{-5pt}
\end{figure*}

Our study was built upon the existing process of our partner ISP, which includes identifying QSnatch infected users and notifying them as shown in \autoref{fig:Method}. Our  mixed-method approach started from the survival analysis of 760 customers for three mutually exclusive categories of malware families: Windows malware, non-persistent IoT malware and persistent IoT malware. These infections were identified and tracked using the daily reports provided by Shadowserver~\cite{Shadowserver2021} which were recorded in the abuse department system of the ISP from the period between May 2020 and May 2021.

Thereafter, we contacted customers infected with QSnatch malware to carry out an interview to understand how they handled the infection. The interview design was informed by the COM-B behavior model \cite{michie2011}, which stresses the importance of individuals' capabilities, motivations, and opportunities to perform a behavior.  The model has been suggested to be applied to understand behavior change in security   \cite{enisa2018cybersecurity}. Finally, the recordings of the interviews were transcribed and coded for its analysis.

\subsection{Survival analysis}

\subsubsection{Determining infections}
As described in the \autoref{sec:background}, our partner ISP connects the IP addresses flagged in the Shadowsever ``Botnet Drone Report'' \cite{Shadowserver2021} with their own customer data to determine which customers are infected. The Drone Report contains data on infections for 112 different malware families, including QSnatch, Mirai and Windows malware. Shadowserver reports are generated daily, and there were no gaps between reports during the period of this study for our partner ISP.

Once infected customers have been identified, a case is created in their incident ticketing system to follow up with customers, and send notifications. A case is closed once the IP addresses belonging to an infected customer are not seen again in subsequent reports. Note that a customer might receive multiple notifications for the same infection if the infection persists for long periods of time. 

In this research, we track users' infections with customers IDs. These IDs are unique and can be associated with multiple IP addresses over time. This way we can track the whole period of infection, even though the IP address of the customer might change in the course of the measurement period.

In the normal workflow, the notifications are sent the day after the Drone Report has reported the infection. However, during the period of this research, the abuse department was transitioning to a new system. This caused delays in sending out some of the notifications. These were randomly distributed across the infections in the Drone Report, thus across all types of malware. The transitioning of the abuse system does not affect the results of this research; instead, this served as a natural experiment. A natural experiment is where a circumstance that was not controlled by the researchers occurs giving the opportunity to evaluate the intervention \cite{leatherdale2019natural}, in this case, the impact of notifications. All infected customers were eventually notified, either the day after the IP address was first detected by Shadowserver or after a second or third detection. 

Our starting point is a set of 760  customers that got notified for  three  categories  of  malware infections:  Windows  malware,  non-persistent  IoT  malware (Mirai) and persistent  IoT  malware (QSnatch). These notifications were sent over the course of a year: May 2020 to May 2021. 
The dataset includes 228 customers infected with QSnatch, 107 customers infected with Mirai, and 425 customers infected with Windows malware. 
For Windows, infections consisted of the following malware families:
Ramnit, Kovter, Citadel, Qrypterrat,  Conficker, Necurs, Sality,  
Caphaw, Downadup, Emotet, Gamarue, Gozi, Necurs, Nivdort, Nymaim, Grypter.rat, Ramnit, Sirefef, Tinba, and Zeroaccess. These Windows malware families have a wide range of capabilities, from banking trojans to worms to ransomware.  We group all the Windows malware families together, rather than comparing individual families. First of all, we are interested in comparing malware categories---persistent/non-persistent and IoT/non-IoT---rather than individual families. Second, for some families the sample size was very small (e.g., 1, 4, 7 or 8 observations). This rules out meaningful comparisons.

\subsubsection{Kaplan-Meier estimates}

We selected all infections with a start date between May 2020 and May 2021. To unravel how cleanup of QSnatch infections compares to other malware categories, we computed the survival time probability for customers infected with QSnatch, Mirai, and  Window malware families. For this purpose, we used Kaplan-Meier (K-M) curves and estimates~\cite{kaplan1958nonparametric,rich2010}. 
To construct the survival time probability and curves of the different malware families, we used the starting date of the infection and the final date of infection from the historical Shadowserver data stored in the incident ticketing system of the ISP (see \autoref{subsec:infection_time} for more details).
As described by \cite{rich2010}, this allowed us to compare all the observations within the groups and begin the analysis at the same point, we check their lifetime until cleanup occurs or the observation period ends. The latter cases are censored. Censoring means the total survival time for the observation cannot be precisely determined since it falls outside the period of data collection \cite{rich2010}. These data points are retained in the analysis, but they are considered as the event did not happen. In this research, observations identified during the last 14 days of the period of observations were right-censored.

The Kaplan-Meier estimator is a non-parametric statistic used to estimate the survival function. The function is defined as: $S(t) = P({X>t})$ \cite{klein2003survival}. In this equation, $S$ is the probability that a random variable $X$, in this case that malware is still on the device, exceeds a specified time $t$. We used the lifelines library \cite{CamDavidson-Pilon2022} to plot the curves and compare them visually and statistically. 

To statistically test whether the differences between survival curves are significant, we use the log-rank test~\cite{bland2004logrank}. This is a method to compare the survival functions of different populations. It compares the estimates of the hazard functions of two groups at each observed event time.

\subsubsection{Infection time} \label{subsec:infection_time}

To construct the survival probability, we needed to estimate the duration of infection. We used a year of historical infection data that the ISP receives from Shadowserver~\cite{Shadowserver2021}, so we consider as ``infection time''  the period between the first time the infection is detected until the last time the infection is seen. For all cases, we had the starting point of an observation, but in some cases, we could not determine if the remediation happened or not since the infection was detected close to the end of the period of observation. Thus, observations identified during the last 14 days of the period of observations were right-censored. 

Note that in the survival analysis of the interviewed users, only 55 observations will be presented. Due to the system transitioning of the abuse department for two customers, we could not retrieve the closing date of the infection. In other words, these two users were notified since they were seen in Shadowserver and added to the incident ticketing system; thus, we contacted them for the interview, but the ISP system did not record the end-time of the infection to calculate the survival probability. 

Note that the infection time as observed in Shadowserver consists of the time it took the ISP to notify the infected customer, the time the customer waited before taking action, the time it took to execute those actions, and the time the infection remained on the device if the actions were unsuccessful in removing the infection.

To avoid any confusion, we should note that during the interviews we asked users if they could roughly estimate how much time they took to perform the remediation steps. We consider this time the users' self-reported time of dealing with the infection. This should not be confused with the total time of the infection, as derived from Shadowserver observations.

\subsection{Interviews}

To understand the process that users follow to perform the steps, and determine if they are able to deal with persistent malware, we developed an interview protocol which we executed in April and May 2021, at the end of the observation period for infections. The interview was a structured interview with closed questions with the opportunity to elaborate on the answer, and some open questions. 

The downside of contacting customers retrospectively is that there was a time difference between the interview and users' actions, which we will discuss more in \autoref{sec:limitations} (Limitations). Also, we chose for Qsnatch-only interviews, rather than a design that would have interviewed people from all three ``treatments" (Qsnatch, Mirai, Windows). This choice has pros and cons. We acquired more data on the challenges of a new and non-studied group, but we cannot compare the answers of the different groups and connect them to the different remediation speeds.

From the total set of customers who suffered a QSnatch infection in the year May 2020--May 2021 (n=228), 45 (20\%) were contacted to carry a pilot to test the protocol (See  \autoref{sec:pilot}). Then the remaining customers, 183 (80\%), were invited to participate in an interview via email. The email stated to customers that they were notified in the past about a QSnatch infection 
and that we wanted to learn about the actions they took, if any, to remediate the infection. Of the 183 customers, 57 (31\%) accepted to participate in the survey. We later checked for selection bias by comparing the remediation rates for the interviewed users versus the non-interviewed users and found no significant difference.

The interview was divided into four parts as described in \autoref{fig:Method}. 
First, we obtained consent from the users to participate in the study as well as recording the interviews, and users were reminded that they could step out at any time. 
Second, we asked if the person we contacted was the one who manages the device, if they received the notification and if they understood the notification. 
Third, different questions about how users handle the infection were asked.
This design was informed by the COM-B behavior change model \cite{michie2011}. More details can be found in \autoref{sec:com-b}, regarding how the principles of COM-B were seen as useful to the study, and how the questions within the interview protocol were based on the COM-B pillars.

Finally, a number of demographic questions were asked, as well as a closing question in which users could add any remark that was not covered during the interview. Next, we thanked the participant for his time, and finished the interview.  
See the complete interview protocol in \refappendix{sec:appendixB}. 
The recordings of the interviews were transcribed and coded using ATLAS.ti software. 

\subsubsection{COM-B and security behaviors} \label{sec:com-b}
The COM-B model has been proposed as applicable in the goal of understanding motivators and blockers for secure user behaviors, both for home users and in organizations \cite{enisa2018cybersecurity}. The pillars of the model (Capability, Opportunity, and Motivation) act as attributes which must all be in place to provide the conditions for a behavior change intervention to be regarded as complete. As we describe in \autoref{sec:remediation_mechanism_manuf_ISP} QSnatch cleanup is complex relative to the number of steps that users have to perform to clean up Mirai \cite{Cetin2019, bouwmeester2021thing, rodriguez2021user} and Windows malware (e.g.running an antivirus). The difficulty of removing a QSnatch infection in users' home networks could be affected by these three pillars. If any of these attributes are not in place, this can translate into a longer time to remove the malware infection.

The COM-B model then stresses the importance of individuals' capabilities, motivations, and opportunities to perform a behavior. These aspects are critical for moving from malware detection to targeted intervention, and ultimately to user's actively adopting and proactively using malware-prevention solutions. Framed this way, the partner ISP was deploying an intervention, to notify users of the QSnatch infection and prompt a new behavior to occur. The COM-B model can help us to understand whether the COM attributes are being supported, and if any one pillar is not sufficiently supported, toward influencing ISP customers to perform a particular behavior. This behavior may or may not lead to the cleaning of infected devices, so we can recommend how the current intervention or future interventions can be improved. To add value to the partner ISP, COM-B is suitable for analyzing customer behavior after they receive the notification, to identify where targeted improvements may be made.

In reference to the COM-B model, we asked our participants a range of questions, addressing various aspects key to a successful behavior; the opportunity presented by the intervention from the ISP, in this case, receiving the notification (including whether it was noticed, and trusted, as in \refappendix{sec:appendixB}); participants' capabilities to parse and act on the content of the notification (such as existing experience with IT systems and if users asked for help), and; if users had any limitations or reservations about performing the steps (such as perceiving a lack of support or tools to complete the steps in the notification, or beliefs about their own capacity or urgency to take personal action). 

\subsubsection{Coding and qualitative data analysis}
Once interviews were completed, they were transcribed and analyzed. Two of the researchers coded the transcripts with ATLAS.ti software using codebook-style Thematic Analysis (TA)~\cite{braun2020one}. Codes were created to label recurring topics, guided by discussion between the two coders to refine the themes. Inter-Rater Reliability (IRR) does not impact the usefulness of emerging themes with this approach, as noted by Braun \& Clarke~\cite{braun2020one} and others~\cite{mcdonald2019reliability}. However, themes were discussed at intervals with the wider co-author team to determine the central themes, where this approach can ensure the reliability of findings \cite{mcdonald2019reliability}. Agreement was reached on seven categories that pointed to core themes in \autoref{sec:interview_results}. The last theme on  \textit{Suggestions} was related to customer feedback, mostly as recommendations for improvements to the service. \textit{Suggestions}, are then included in the Results section (\autoref{sec:results}) where they relate to other core themes and not as a stand-alone subsection, more specifically in  \autoref{sec:commnication_channel}, and they were also shared with the partner ISP after concluding the research, to inform considerations for improvement to the support that the ISP gives to its customers (See \autoref{sec:ethics}).
 
\autoref{tab:atlas_ti} shows an overview of the core themes, along with examples of codes within each core theme, and the percentage of respondents that discussed those themes as an indicator of the prevalence of each theme across the participant cohort.

% \usepackage{booktabs}
\begin{table*}[!ht]
\caption{Summary of qualitative coding scheme}
	\centering

\begin{tabular}{@{}lllll@{}}
% \toprule
                     \textbf{Themes}            & \textbf{Code examples}                                                                                                                                                                                                                                                                                                                                                         & \textbf{\begin{tabular}[c]{@{}l@{}}Respondents   \\ n=57\end{tabular}} &  &  \\ \midrule[2px]
Receiving and understanding the   notification & Receiving notification, understanding notification message                                                                                                                                                                                                                                                                                    & 57 (100\%)                                                         &  &  \\

Cleanup effort                                 & Cleanup time, time to execute steps                                                                                                                                                                                                                                                                                                                                                   & 49 (86\%)                                                          &  &  \\
Technical (security) ability                              & IT profession, IT experience                                                                                                                                                                                                                                                                                                                                                    & 56 (98\%)                                  &  &  \\

Beliefs about risk                             & Consequences of not executing steps                  & 54 (95\%)                                   

&  &  \\
Responsibility                                 & Personal responsibility, ownership                   &53 (93\%) &&\\                                                                                                                                                                       Communication channel                                          & Trust, distrust                                                                                                                                                       & 41 (72\%)                                                                                                                         &  &  \\
Suggestions                                & Suggestions to the ISP, comments   to the ISP, congratulations  & 34 (60\%)                                                          &  &  \\ \bottomrule[1px]

\end{tabular}
\label{tab:atlas_ti}
\end{table*}

 \subsubsection{Pilot interviews} \label{sec:pilot}
It was important to arrive at a robust study protocol, not only for engaging with real-world users outside of a controlled laboratory setting, but also with participants who were customers of our partner ISP. To test the interview protocol, 45 customers were contacted, 14 customers did not answer the call, 14 opted out, 3 numbers were out of service, and 14 customers decided to participate in the research. From the 14 customers who participated, 7 (50\%) customers were showing up as remediated at the moment we talked to them and 7 (50\%) were showing up as still infected. The main change after the pilot was to ask users if the device was used for private or business purposes or both. We uncovered that some customers use their devices for these different purposes. The pilot interviews led us to decide to have more precise questions and less open questions. This was based on the willingness of ISP customers to participate in the pilot since we learned that customers would not spend on average more than fifteen minutes engaging with the data collection, this without including the time that the researcher carrying out the interview took to introduce himself, describe the research, and gain consent from the participant. These 14 pilots interviews are not included in the dataset of 57 interviews that forms the basis of the interview study.

\subsubsection{Interviewed participants}
After completing the pilot, we conducted 57 interviews. The age of these customers ranged from 22 to 63 years old. Four (7\%) participants self-report their gender as female, and 53 (93\%) as male. Most participants, 46 out of 57, used the QNAP device for private purposes, 5 used the device for business purposes and 6 used the device for both business and private purposes. No incentive was provided to participate in the research.

\subsection{Cleanup time after notification}
\label{sec:cleanup_time_notification}
While we derive the infection time (or infection duration) from the Shadowserver data (\autoref{subsec:infection_time}) recorded in the incident ticketing system of the ISP, we also want to know how long it took users to clean up the infection after they were notified. 
In this research, we defined as ``clean"  a user device which stops showing up as infected after being notified. On the other hand, if the observation continues showing up in the feed, we considered the user as ``not clean". 

Retrieving the time stamps of the notification(s) was a labor-intensive manual process. Since the abuse department was transitioning to a new system, it was required to manually check the IDs for a period of a year and be careful about not missing notification. Thus, we were only able to do this for the interviewed users, except for six customers, where the abovementioned system transition meant we could not retrieve this data. 

In the end, we collected the notification time stamps for 51 users. For this group, we could determine when the cleanup happened in relation to the notifications received by the customer. Unfortunately, we could not compare these findings with the Mirai and Windows infection groups.

\vspace{-5pt}

%% file: ethics.tex
% !TeX root = paper.tex
\subsection{Ethical considerations}\label{sec:ethics}

The human research ethics committee  of our institution approved the interview protocol of this study (Reference number: 1490). Consent for anonymously taking part in this research, as well as for recording the calls, was obtained from the participants. They were also reminded that they could stop the study at any time. 

Following the Menlo Report~\cite{Dittrich2012}, we were guided by the ethical principles of respecting people, respecting the law, justice, and beneficence. 
Regarding respecting people and law, we followed all the guidelines and privacy policies of our partner Internet Service Provider, and personal data never left the Internet Service Provider's premises. One author was embedded in the ISP, and in consultation with the ISP’s privacy team and within terms of service linked user IPs and interviewees then produced an anonymized dataset used in our data analysis. Further, as pilot participants stated that they had limited time to participate in research, the protocol for the main study was adapted to respect this.

Regarding justice, the study did not benefit specific groups over others. All infected customers were contacted for the study and had equal opportunity to share their experiences and provide feedback.

Regarding beneficence, we did not interfere with the ISP's beneficence and all subscribers affected by QSnatch malware were notified of the infection, so they were able to protect themselves and others from this threat. The goal of the interviews was to learn how users experienced the remediation process to improve the support that the ISP can give for its customers. Also, our research aims to understand how users deal with persistent IoT malware in order to benefit society at large.

%% file: results.tex
%\section{Results}
\section{Results}
\label{sec:results}

\subsection{Survival analysis}

In this section, we answer the question of whether persistent IoT malware, namely QSnatch, is more difficult to remediate compared to persistent Windows malware and non-persistent IoT malware, namely Mirai. Higher difficulty would result in longer infection times.

\autoref{tab:summary_cleanup} shows the cleanup success and the infection times (mean, standard deviation and the distribution) for each of the three categories of malware families, namely Windows malware (n=425), Mirai (n=107), and QSnatch (n=228).

\begin{table*}[t!]
	\centering
	%\scriptsize
	\caption{Summary statistics per group of infection type, with remediation outcomes.}
	\label{tab:summary_cleanup}
	\begin{tabular}{l r r r r r r r r r}
	&&&\multicolumn{7}{c}{Infection time (days)} \\ \cmidrule[2px]{4-10}
	\multicolumn{1}{c}{Group} & \begin{tabular}[c]{@{}c@{}}Sample \\ Size\end{tabular} & \% clean & \multicolumn{1}{c}{\begin{tabular}[c]{@{}c@{}}Mean \end{tabular}} &\multicolumn{1}{c}{\begin{tabular}[c]{@{}c@{}}Standard\\deviation\end{tabular}}
	&Min&25\%&50\%&75\%&Max\\ 
		\midrule[2px]
		
		Windows malware &  425 & 97\% & 36 & 76  & 0  &0  &0 &26 &359  \\
    	Mirai &  107  & 100\% & 19 & 36  &0 &0 &1 &24  &182  \\
		QSnatch &  228 & 91\% & 108 & 110 &0 &3 &76 &181 &365 \\
		\bottomrule
	\end{tabular}
	\vspace{0.5cm}
\end{table*}

\begin{table*}[t!]
	\centering
	
\begin{threeparttable}

	\caption{Summary statistics for interviewed and non-interviewed groups exhibiting QSnatch device infections.}

	\label{tab:summary_int_noint}
	\begin{tabular}{l r r r r r r r r r}
	&&&\multicolumn{7}{c}{Infection time (days)} \\ \cmidrule[2px]{4-10} 
	\multicolumn{1}{c}{Group} & \begin{tabular}[c]{@{}c@{}}Sample \\ Size\end{tabular} & \% clean & \multicolumn{1}{c}{\begin{tabular}[c]{@{}c@{}}Mean \end{tabular}} &\multicolumn{1}{c}{\begin{tabular}[c]{@{}c@{}}Standard\\deviation\end{tabular}} &Min&25\%&50\%&75\%&Max\\ 
		\midrule[2px]
		QSnatch -- Not interviewed &  173 & 89\% & 112 & 116  &0 &3 &76 &181 &365 \\
    	QSnatch -- Interviewed &  55* & 100\% & 94 & 86  & 0 &2 &76  &157  &273 \\
		
		\bottomrule
	\scriptsize

	\end{tabular}
	  \begin{tablenotes}
      \scriptsize
      \item * Note that the interviewed group is n=57. We could  not retrieve the infection end dates for two users, due to the system transitioning at the abuse department (See \autoref{subsec:infection_time}), thus in this table n=55 for the interviewed group. 

    \end{tablenotes}
  \end{threeparttable}
\end{table*}

The mean infection time of Windows malware is 36~days with a standard deviation of 76 days, the median infection time is 0 days and the maximum infection time is 359 days. For Mirai, the mean infection time is 19 days, with a standard deviation of 76 days, the median infection time is 1 day, and the maximum infection time is 182 days.

In contrast, for QSnatch the mean infection time was much longer than the other two malware categories: 108 days, with a standard deviation of 110 days. The median infection time  is 76 days, and the maximum infection time is 365 days. 

The mean infection time of QSnatch infections is three times higher than the mean time for Windows infections and five times higher than the mean time for Mirai infections. 

\begin{figure}[t!]
	\centering
	\includegraphics[width=1\columnwidth]{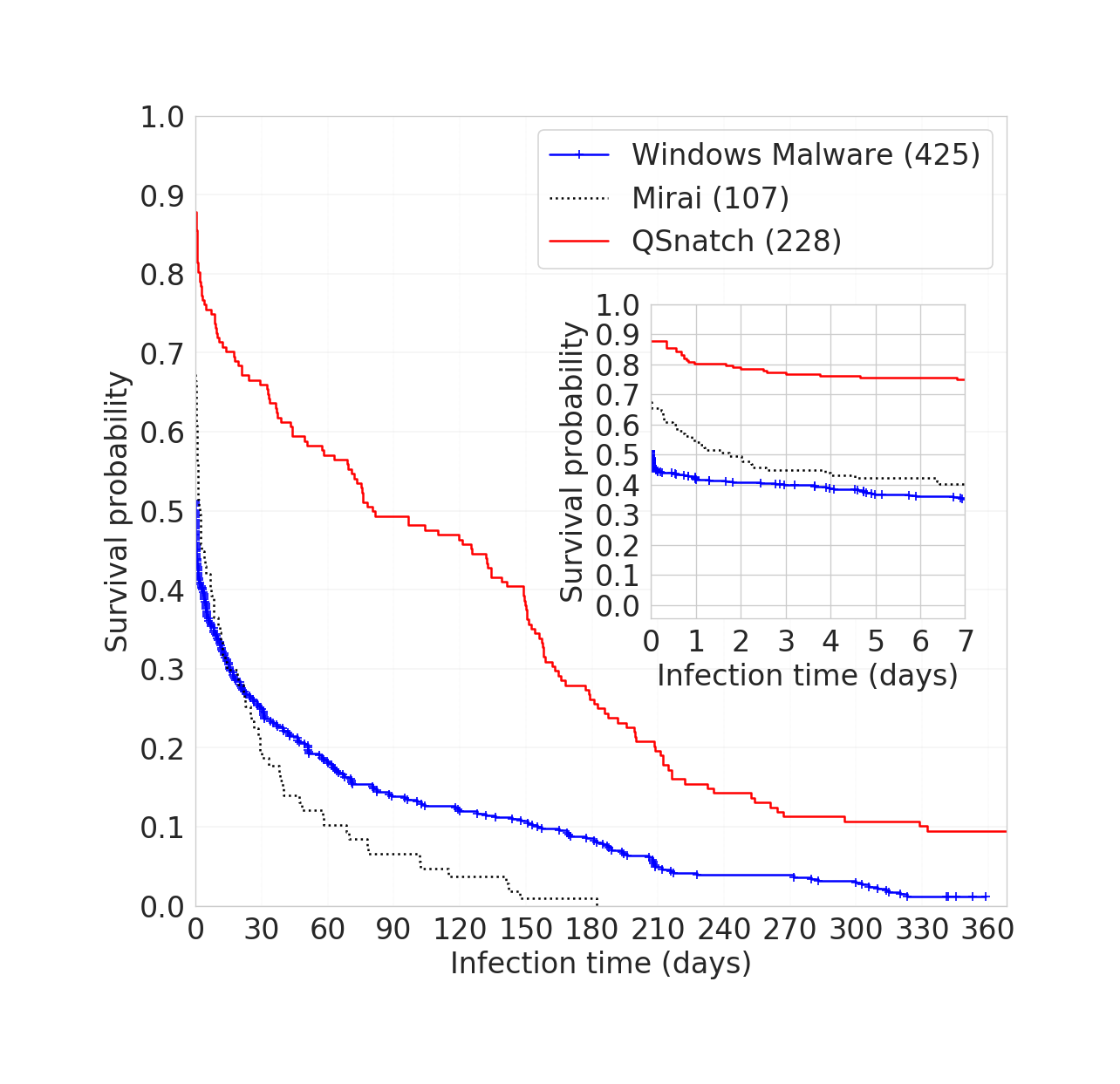}
	\vspace*{-10mm}
	\caption{Survival probability QSnatch vs Window malware vs Mirai 
}
	\label{fig:survival}
\end{figure}

For a more comprehensive analysis of the data, we computed the survival probability for each malware category using Kaplan-Meier estimates. \autoref{fig:survival} shows that after 180 days, around 30\% of the QSnatch infections are still alive, while only 10\% of the Windows infections remain, and none of the Mirai infections. 
\autoref{fig:survival} inset figure also shows that within 7 days after the infection QSnatch remain stable at almost 80\%, while Mirai and Windows malware already drop to almost 50\% or lower.

Consistent with \cite{Cetin2018}, we have also observed a high cleanup rate at the beginning of the infection time for Windows malware and Mirai, even though in our study, most participants were notified via email rather than put in a quarantine network. We do not observe this same pattern for QSnatch.

The log-rank test reports whether there is a
significant difference between the QSnatch and the two other groups. We find that the differences with both groups are highly significant: Mirai versus QSnatch (log-rank test: ${X^2}$= 96.22 with  $p=0.00$) and Windows malware versus QSnatch (log-rank test: ${X^2}$= 80.27 with  $p=0.00$).

To check whether our interview study suffered from selection bias, where the people who were willing to participate might also be more committed to conducting remediation, we analyzed the infection time data for both groups, interviewees as well as non-interviewees.

From the total users infected with QSnatch in the period of observation (n=228), we interviewed 57 users. We need to remind the reader that we could not retrieve data of 2 participants for the survival curve,  thus the number of observations in the graph is 55 (See \autoref{subsec:infection_time}).

\autoref{tab:summary_int_noint} shows the summary statistics of the  QSnatch not-interviewed users versus the interviewed users. The mean infection time of not-interviewed users is 112 days with a standard deviation of 116 days. For the interviewed group, the mean infection time is a bit lower, 94 days, with a standard deviation of 86 days. 

\begin{figure}[t!]
	\centering
	\includegraphics[width=1\columnwidth]{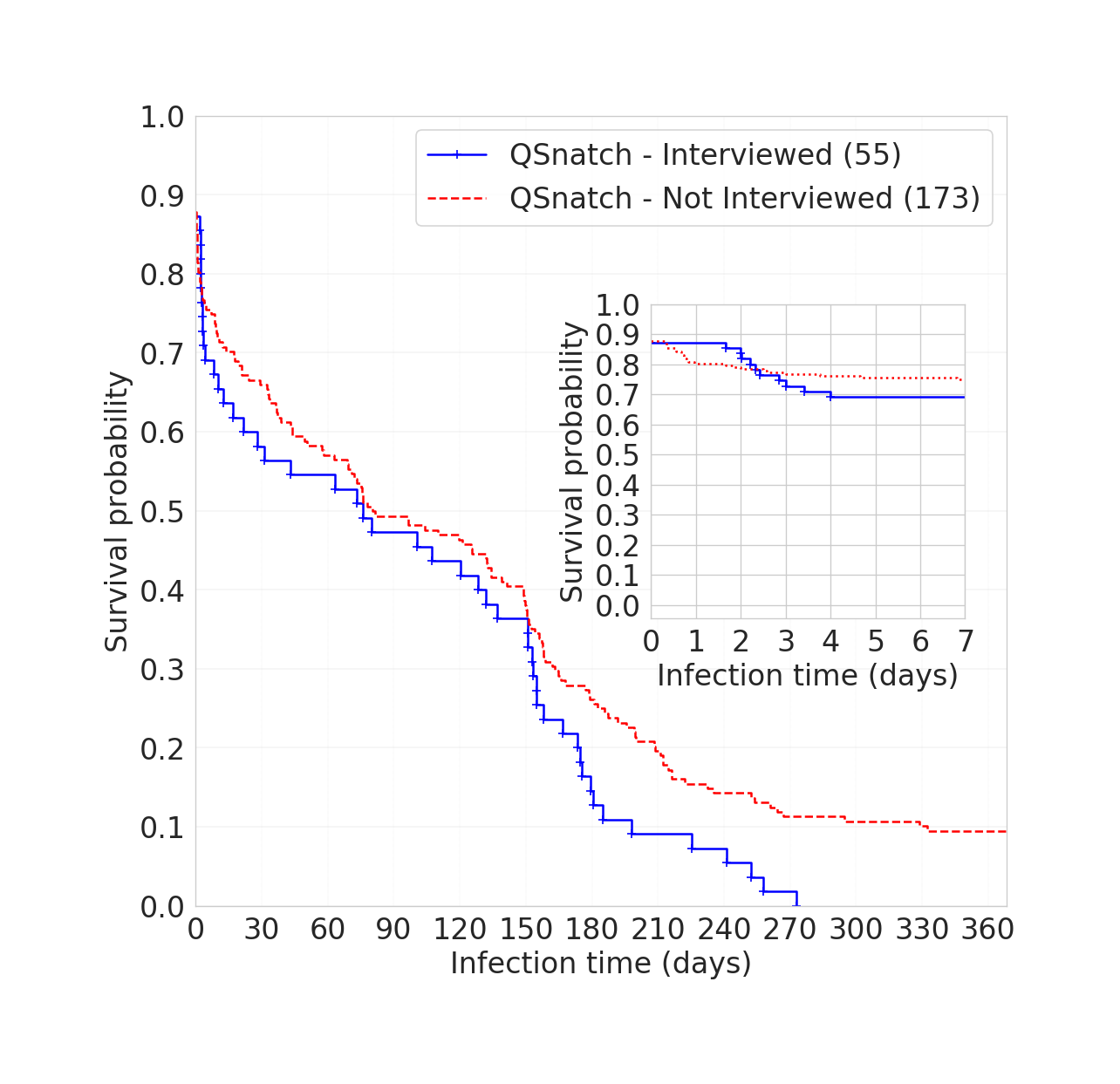}
	\vspace*{-10mm}
	\caption{Survival probability participants vs not interviewed users. 
}
	\label{fig:survival_interviewed}
\end{figure}

\autoref{fig:survival_interviewed} shows the survival probability of both groups. They are very similar. Only at the tail end of the plot do we see that 10\% of the non-interviewed group remains infected at the end of the period, while all of the interviewees have remediated. 
We did a log-rank test to check if there were differences between the groups. The log-rank test reports no statistical differences between the groups at a 5\% significance level (${X^2}$= 3.09 with $p=0.08$).

\subsection{Interviews} \label{sec:interview_results}

Given that participants have to comply with many different steps, as described in
\autoref{sec:methodology}, different questions were asked to understand how they handled the remediation process.
The interviews were transcribed and coded, different themes emerged that will be described in this section. We focus on the most prominent themes which emerged from the interview analysis.

\subsubsection{Receiving and understanding the notification}

We asked participants if they recall receiving the notification and recall performing the recommended steps. In total, 53 (92\%) recall receiving the notification, and only 4 (8\%) participants either said they did not receive it or were unsure.  
\autoref{tab:notification} shows a summary of the participant's answers.
The majority of participants 44 (77\%) recalled doing the steps. Of the remaining 23\%, most customers reported charting their own course to solve the infection. Five (9\%) participants reported updating the device, while three (5\%) reported following the manufacturer's steps, and four (7\%) turned off the device. One (2\%) respondent reported calling the QNAP helpdesk to solve the infection. The four participants who said that they did not receive the notification or were not sure of receiving it, were among the participants that mentioned doing some of these different steps.

\begin{table*}[!ht]
\vspace{-5pt}
	%\scriptsize
	\centering
	\caption{Recall of responses to ISP notifications for interviewed participants (N = 57). }
	\label{tab:notification}
	\vspace{-5pt}
	\renewcommand{\arraystretch}{1.2}

	\begin{tabular}{l l l l}

% 	\midrule
	\textbf{Recall of responses} & \textbf{Interviewee response} & \textbf{No. Interviewees} \\
	\midrule[2px]

	\textbf{Recall receiving notification} 
		& Yes &     53 (92\%)  \\
	 & No  & 2 (4\%)  \\
		& Not sure & 2 (4\%)  \\

	\textbf{Recall performing steps} 
		& Yes &     44 (77\%)  \\
		& Update & 5 (9\%)\\
		& Turn off device &4 (7\%) \\
		& QNAP steps & 3 (5\%)\\
		
		& Call QNAP helpdesk to cleanup & 1 (2\%)\\

	\bottomrule[1px]
	\end{tabular}
% 	\end{adjustbox}
\end{table*}

\subsubsection{Cleanup effort}

We asked the participants to estimate the time they had spent on their remediation actions. There was high variability in the reported times and four categories emerged.

\autoref{tab:self-reported_time} shows the self-reported time participants invested following the steps.
The largest group of participants, 25 (44\%), gave answers in the range of more than 15 minutes up to 1 hour. Ten (18\%) participants reported taking up to 15 minutes. Eleven (19\%) reported answers that ranged from more than one hour up to twelve hours. For example, P45 reported \textit{``The steps you indicated were completed quickly. That would have taken half an hour. But what actually should happen was that it took half a day of work to solve it completely."} Finally, three (5 \%) participants reported taking more than 12 hours up to 24 hours. In cases where participants reported almost a day of work, they did not refer to the actions themselves consuming so much time, but that their overall remediation process took that long. The NAS took time to execute various instructions as well. For instance, P47 stated \textit{``I think (I spent) an hour per NAS myself, but the device can easily be working for an entire day''}. Eight (14\%) participants did not answer the question.
All in all, participants reported being able to execute the actions swiftly or at least within a day. 

\begin{table}[!t]
	%\scriptsize
	\centering
	\caption{Self-reported time invested in remediation actions }
	\label{tab:self-reported_time}
	\renewcommand{\arraystretch}{1}
	\begin{tabular}{ l c }
		\textbf{Time} & \textbf{No. Interviewees} \\ 
		\midrule[2px]
		Up to 15 min&  10 (18\%) \\
		More than 15 min, up to 1 hr & 25 (44\%) \\
	    More than 1 hr, up to 12 hrs  & 11 (19\%) \\ 
		More than 12 hrs, up to 24 hrs & 3  (5\%) \\ 
		No answer &  8 (14\%) \\ 
		 
	    \bottomrule[1px]
		\end{tabular}%
	
\end{table}

Next, we looked at the cleanup time: the time between the first notification and the end of the infection. We have the time stamps of all notifications for 51 participants. (As explained in \autoref{sec:cleanup_time_notification}, we could not retrieve this information for six customers.) 

The largest set of participants, 24 of 51 (47\%), acted after the first notification. All of them cleaned up in one or two days after the notification. 
Note that some of these participants received their first and only notification very late, because of the random delays caused by the abuse system transition. If they were not immediately notified upon the first observation in the Shadowserver data, some time would pass before they are observed again in the Shadowserver data. In some cases, even this second or third observation did not trigger a notification. This meant that their infection time could be very long. 
Because of the random delays, nine of these customers have a total infection time between 31 days and 241 days.
Yet, once the customer is notified, the remediation takes place within two days at most. 

These random delays unintentionally tested the effect of the notification. Before the notification, those users were infected with QSnatch for one or more months. It underlines the necessity of the ISP notification. Apparently, there is no alternative path towards remediation.

Next, there is a group of customers who did not act shortly after the first notification. The number of notifications that a customer received is clearly connected with the overall duration of the infection.
21 of 51 participants (41\%) received between 2 and 9 notifications. For these customers, the duration of infection ranged between 8 days to 252 days. 
Finally, there were 6 of the 51 (11\%) participants who were notified between 11 and 18 times. The duration of the infection was between 128 and 273 days.

From the interviews, we could identify reasons why participants did not act immediately on the notifications.
We found evidence of users planning the remediation tasks consistent with \cite{van2000procrastination} that might have delayed the action. 
P15 said he wanted to wait until he \textit{``could take my time on a Sunday to try to solve this”}. Another interviewee, P8, referred to outsourcing the tasks. He hired an IT provider to do the steps and that process took a while. This participant received 11 notifications before finally showing up as cleaned. P19 said he received the notifications while being out of the country and without remote access to the device, so he had to wait until he got back. This participant received 11 notifications. Other participants said they did not see the email immediately, because it arrived in a mailbox that they do not frequently check. P52: \textit{``The emails arrived early with me, but in a mailbox that I barely ever read. That is why I checked this only very late and took care of the situation.”}. This participant received 13 notifications. P20: \textit{``It took a while before I saw [the notification]. Once I saw it, I took action”}. This participant received 18 notifications from the ISP.  

P50 was one of the customers with the most notifications, 18 in total. When he was asked whether he found the steps useful. He said: \textit{``Yes, although it is difficult to know whether it is useful. Initially, I ignored the email twice or so, because I wondered whether this was officially from the ISP. You get so many emails these days that you aren’t sure. But after receiving it repeatedly, 2 or 3 times, I thought: OK, this is serious, let’s take action."}

The more participants overlooked the notifications, the longer their infection time. That said, the recurring notifications did at some point spur them into action.
Only one customer cleaned up long after the last notification. 

Our data shows that many participants acted on or close to the first notification they received. It is also important to note that email was an effective channel through which to reach many users, similar to \cite{cetin2019tell}. Our findings demonstrate that email can be a cheap and scalable alternative to prompt participants to remediate, compared to walled gardens, letters or other notification mechanisms~\cite{263802}. 
However, for some users an alternative notification mechanism, rather than a repeat notification, might be needed.

Finally, the random delays in the first notification also demonstrated how important the notification process is for persistent malware.

\subsubsection{Technical ability} 

The remediation process entails several relatively complicated steps, compared to the remediation advice for Windows and Mirai infections \cite{Cetin2019,rodriguez2021user,bouwmeester2021thing}. Yet, most participants report needing only a short time to conduct the steps. In line with this, we also encountered very little evidence that participants felt the steps were difficult to execute. Only one person mentioned any doubt as to how to perform to remediate the problem. Most participants described it as a straightforward task. This sounds a bit paradoxical: the task is relatively difficult, the infection took long, users were not aware of the infection until the ISP notified them, yet very few users expressed experiencing any difficulty.

\begin{table}[t]
% 	\scriptsize
	\centering
	\caption{Self-reported IT experience}
	\label{tab:IT_experience}
	\vspace{-5pt}
	\renewcommand{\arraystretch}{1}
	\begin{tabular}{ l c }
		\textbf{Type of experience} & \textbf{No. Interviewees} \\ 
		\midrule[2px]
		 IT professional     &32 (56\%) \\
	    
		No experience with managing IT& 15 (26\%) \\

	    Some experience with managing IT &9 (16\%) \\

		No answer&  1 (2\%) \\
	    \bottomrule[1px]
		\end{tabular}%
		
\end{table}

This paradox points to their skill level related to capability in the COM pillars. We asked interviewees about their IT experience. 
\autoref{tab:IT_experience} summarizes the answers across three main categories.
A stunning 56\% of the interviewees said they were IT professionals. 
P43 said:\textit{ ``I've worked in IT for 20 years"}. 
Several participants said they worked as system administrators. For example, P44 mentioned: \textit{``I'm kind of a system administrator at work. I'm fairly well versed in it"}. 
Others work in software development and programming---for instance, P56:\textit{ ``I have my network in my home, you know, I can use it, this is also my job, my profession is
programming"}. Some reported working in network security and automation. 

A second group, 16\% of the participants, claimed some experience managing IT, though generally out of interest rather than in a professional capacity.

To illustrate, P52 said that \textit{``I happen to work at [ISP NAME] myself"}. Others mentioned working on their own networks at home as hobbies, managing their own servers, and similar activities. P28 reported: \textit{``So I've always had a server running in my own network for 15 years. A hobby that got out of hand."} 
Only 15 out of 57 interviewees (26\%) said they did not have any real experience with IT. Four (27\%) of these customers reported looking for help from IT professionals, an acquaintance with IT experience, or a friend. It is worth noting that from the users who did report some IT experience, one asked for help as well, but the help he referred to was contacting the QNAP helpdesk.

Our findings suggest a self-selection process or early adopters at work \cite{dedehayir2020innovators}. NAS devices attract a user population that is significantly more skilled than the average user population. This would explain why the participants handling the remediation had some tolerance to execute the complex process. In fact, for an IT professional, the frame of reference is different. They are more likely to compare the QSnatch remediation actions to IT admin tasks, rather than to the consumer tasks of running an AV tool on a Windows machine or changing the password on a Mirai-infected IoT device. In that light, the QSnatch remediation process is not particularly difficult. Clearly, this finding is unlikely to hold for other IoT devices that are more widely distributed among consumers. 

Interestingly, only a few participants questioned how the ISP knew about an infection that they did not know about, or about how they got infected given that they have self-reported IT experience. Meaning that they did not question their setup or how they got infected. P14 stated: \textit{``I'm curious
how that [infection] came about. I have to say that I thought it was strange because I suspected that I had
nothing open. I had all those services turned off. I only use it as a local NAS in my local network. So all ports to the outside were turned off. And that makes it very strange that that is possible."}. In total only five participants were doubting how the infection happened or how the ISP knew about their infection.

\subsubsection{Beliefs about risk}\label{sec:beliefs}

We asked participants what they believed would happen if someone were not to follow the recommended steps. Participants expressed certain beliefs which contribute to their decisions about whether to act upon the notification. We found a variety of beliefs about viruses, comparable to those identified in other work examining home participants' mental models of security~\cite{wash2010folk}. 

Most participants state that if the steps are not followed, malicious activity may be directed toward them. For instance, the malware stays, data is lost or held for ransom, or the device becomes accessible to attackers, among other beliefs.
For example, 18 (31\%), stated that unless action is taken, the malware will stay on the device. Two of these participants added that this could bring consequences to their network safety, and two participants mentioned that this could affect others. One participant said that the malware could spread. 13 out of the 57 participants described data loss or theft as an anticipated consequence of not completing the steps; one of this same group also mentioned the possibility of a Distributed Denial of Service attack. For instance, P16 expressed: \textit{``it may just be that they can access your photos, for example, and do something with them, ransom and so on''}. P29 stated that \textit{``then it [the malware] releases files that may be private''}.

Six other participants (10\%) described that a compromise of the data on the device could be possible or that the device is made openly accessible for attackers and exploits.
Three participants believed that they would not have access to the device due to malware (which potentially contradicts their having use of it at the time). Three participants mentioned that they could lose their Internet connection. This can be associated with the fact that the ISP notification stated that if the respondent did not complete the steps, that there was then the possibility of temporarily placing their connection in quarantine.  Three also stated varied beliefs like the ISP would get into problems, that they would get into problems for many years, or that not doing the steps was not an option. 
Three participants were unsure of what could happen.

Most of the expressed beliefs,  similar to \cite{Fagan2016}, were about how participants think the malware would affect them individually rather than thinking about how the infection could affect others.

Relating to the clean-up behaviors, we then see that our participants were completing the steps and motivated to do so. This demonstrates a close link between security beliefs and protective behaviors \cite{wash2015too}; where Wash \& Rader found that individuals with a strong belief that viruses caused problems then self-reported taking action to protect themselves, we have real-world evidence here of this being borne out for consumer IoT devices (independent of the accuracy of the belief).

\subsubsection{Responsibility}

When asked, 53 out of 57 interviewees (93\%) said they felt responsible for cleaning up the device.
Most of them, 34, expressed that the device belongs to them, they manage it, it is in their own network, and they felt responsible for solving security issues. To illustrate, P1, stated: \textit{``Yes (I am responsible), my children my wife use the
NAS so it must all be safe and there are also so private
things stored there also, like tax data''}. 
Five participants connect feeling responsible to being informed of the problem via the notification. To give an idea, P48 said: \textit{``yes, (I am responsible) because I was asked and I manage that system at home. So then I am responsible for those steps"}.
The rest of the participants expressed diverse reasons why they felt responsible for doing the steps, either they indicated that they were the ones having the problem, that no one else would do it, or that they felt responsible because they wanted to get rid of the malware before it caused potential damage.

Beliefs about responsibility are important, as other research has found that individuals may otherwise defer or delegate responsibility to other people \cite{dourish2004security}. We did not see this with the majority of our participants, aside from the few who approached an outside IT specialist for help. Even this action can be seen as a form of taking responsibility.

Haney et al. \cite{haney2021s} asked an open version of this same question to smart device owners, finding a mix of perceptions across personal, manufacturer, and government responsibility; the majority of their participants stated at least partial personal responsibility for the security of their devices. Interestingly, their participants focused on personal responsibility specifically around fixing lapses or precautionary measures around device security which may result in exposure to risks -- this tallies with the setting of our study, where our participants are uniquely queried about real-world infections of their own smart devices, and expressed personal responsibility to resolve the issue.

\subsubsection{Communication channel} \label{sec:commnication_channel}

During the interview, participants were offered the opportunity to discuss or mention things that they considered important that were not asked by the interviewer. Some participants discussed the trust issues they had with the notification. In total 12 (21\%) participants mentioned feeling some distrust towards the ISP notification. Where participants provided customer feedback, most of their suggestions about the service related to the communication channel.

P27 stated: \textit{``Those messages from [ISP NAME] looked very much like it was all fake, so to speak. So I was a little unsure about that too."}. Also, P33 mentioned \textit{``The mail I received from [ISP NAME], I got it in the spam folder, so I almost deleted it. [...] I liked it, I think it's a very nice initiative from [ISP NAME], but to say that it is very normal, no. So it would almost look like someone is trying to trick me about my device. So the ISP should communicate a little better about that."} 
Several users recommended to make the communication more trustworthy. 

The level of distrust is higher than reported in a previous study, where only two users distrusted the notification via email \cite{Cetin2019}. The higher level of distrust might reflect the technical ability of the NAS owners, compared to the broader user population in the earlier study.
Another explanation could be that participants received the notification during the COVID-19 pandemic. They were working from home and might have been more careful with the emails they received. Consistent with \cite{Stock2018}, the trust issues around the email could have played a role in delaying the actions as well.

%% file: related_work.tex
% !TeX root = paper.tex

\section{Related Work}\label{sec:related_work}

Before the past two decades, Windows malware has occupied the security community \cite{cozzi2018}. Also, some Windows malware, such as Conficker, remained in users' machines for many years \cite{Asghari2015a}. With the proliferation of IoT devices, now attackers are shifting to IoT persistent malware \cite{brierley2020persistence,bock2019assessing,Fbot,Matryosh} since these devices have several advantages for attackers, like low computational capacity \cite{Kolias2017}, thus they cannot count on protections such as antivirus which Windows systems do. The current state of the art has also learned about Mirai, a non-persistent IoT malware, that can be removed by rebooting the device and changing passwords \cite{Antonakakis2017,Cetin2019, bouwmeester2021thing, rodriguez2021user,estalenx}, however, in this research we dealt with QSnatch, a malware, that needs convoluted steps from users to be remediated, and does not count on the same mechanisms for removal from Windows malware or Mirai.

Users were notified about a QSnatch malware infection in their home networks.  Li et al. \cite{Li2016} studied notification content and mechanisms in terms of webmasters cleaning up compromised servers. They observed that contacting the webmasters directly increased the likelihood of cleanup by over 50\%. In this study, we contacted the person who managed the network access storage device, and we observed that 45 (78\%) of the participants did the recommended steps, and 13 (22\%) charted their own course to solve the security problem. 

Stock et al. \cite{Stock2018} and Cetin et al. \cite{cetin2019tell} sent notifications to vulnerable domains and described low remediation rates. They highlighted the limitations of email notifications and the breach between taking action and knowing about the problem.  In our work all customers were notified via email only, and for some users the first email notification was enough to take action. However, some participants needed multiple notifications to act.

Li et al. \cite{Li2016a} notified network operators about security issues in their networks revealing that different notifications have different outcomes, but in general notifications have a positive impact on remediation. Dumeric et al. \cite{Durumeric2014} sent notifications for vulnerable  Heartbleed servers and found a beneficial influence in patching. Cetin et al. \cite{Cetin2018} found high remediation rates for Windows-based malware cleanup.  In this research, we observed total cleanup after participants being notified. This could be explained due to the capability that most users self-report.

Vasek et al. \cite{179521} studied how detailed notifications caused more remediation of compromised websites than short notifications. In this study, we found that users benefited from a tailor-made precise advice to execute the steps to solve QSnatch infection. 

Different work on IoT malware notifications  \cite{Cetin2019,bouwmeester2021thing, rodriguez2021user} highlight that once users are aware of an IoT malware infection, they are motivated, comply with the steps and cleanup. Our findings demonstrate that even with more convoluted steps users put time, effort and take responsibility to remediate the infection.

\vspace{-5pt}

%% file: discussion.tex
\section{Discussion} \label{sec:discussion}

Our observations of network data illustrated that the mean  infection  time  of  persistent  IoT  malware  is greater than that of Windows malware and memory-resident  IoT malware; QSnatch infections may persist for several months, as also shown in our data. In terms of successes, we have found real-world  evidence  of our participants successfully mitigating persistent IoT malware. This demonstrates a close relationship with the intervention of the participants' ISP, where the QSnatch-infected devices of the customers we interviewed were remediated at a time close to having received a notification from the ISP. Issues arose in noticing one notification in a series of notifications, as the prompt to take action, and in subsequently planning to take action. In this section we discuss the wider implications of our quantitative and qualitative results.

\subsection{Success and timeliness of remediation}

The participants we interviewed as part of this study all reported taking action to remediate; all were seen to no longer appear in the infection data shortly after receiving a notification. This implies that at least for participants such as ours, who believe they comprehend and can action advice when prompted, that this model of ISP notification is successful. Many participants were thankful for the notification. 

No participant acted prior to receiving a notification, even if their infection was already going on for months. They did not report acting on unexpected device behavior before receiving the notification, as might happen with malware that is generally used to target others outside of the network. Given the proximity of a notification to remediation for participants, we posit that they may well have not taken action if they had not been notified. Natural remediation did not occur either (as has been noted can occur for non-persistent malware infections such as Mirai) \cite{Cetin2019}.

We see from our results that, generally, those participants who took longer to remediate had received more notifications before eventually acting on one. It is less a question of whether we need to help people to successfully remediate, and starts becoming a question of whether we want them to remediate \textit{sooner}. For researchers, this highlights the importance of combining self-reports with technical data, to understand where users are not noticing notifications compared to what they report \cite{redmiles2018asking}. This includes any contributing circumstances, such as seeing the notification when not being near the affected devices and being able to act on the advice (and forgetting it shortly after). In studying operating system warnings on personal computers, Krol et al. \cite{krol2012don} found that over 80\% of their participants were observed to ignore one or other warning, more than those with higher computing proficiency. Egelman et al. \cite{egelman2008you} found participants receiving a passive phishing warning mostly seemed to ignore it, as compared to active warnings which require explicit interaction -- email notifications follow a similar format. 

\subsection{Learning from the idealized IoT user}

In a way, our study found an idealized version of ISP-managed remediation -- the ISP has done what is within their power (acquire abuse data and send a notification) and our participants, for the most part, have received the notification, understood it, acted on it, and then their network is seen as being cleaned. There are, as mentioned above, some inconsistencies in that story, foremost that some participants required many notifications before acting (Opportunity).

This user population arguably consists of `early adopters' of what is currently a niche device (network access storage devices), whose response to this emerging threat could inform what we can reasonably expect of users of varying expertise as this family of devices sees more widespread use. Foremost, this user group was relatively `cheap' to help -- they were told what to do (Opportunity), they did it (Motivation), and it worked. We cannot assume this would hold for other groups of smart device users, especially those with less technical experience (Capability).

The role of personal responsibility in keeping IoT devices secure has been highlighted in other work \cite{haney2021s}, and further, that suitably informed personal responsibility requires understanding of communicated risks, the opportunity to act, and to know how to act. We saw a few participants take independent action to verify the right steps to take, rather than follow the notification steps exactly (as in \refappendix{sec:appendixA}). This suggests that less tech-savvy users may also need advice pitched at a suitable level of competence -- a few of our successfully-remediated participants reported updating the device, or calling the manufacturer for support, which are both approaches which can be adapted to less technically-experienced device owners as they rely less on an assumption of personal technical ability.

\subsection{Self-efficacy and device compromise}

Our participants seemed confident of their capacity to clean up the devices. Despite having been informed that their devices were compromised, none mentioned being surprised or doubting the correctness of their device setup. However, five participants did enquire as to how the ISP knew about the infection and, in a manner, questioning whether there was an infection. It may be that our experienced participants do not associate remediation efficacy with device setup efficacy. Otherwise, the IT-related work that many were involved in may have desensitized them to device infections being an issue, especially if they perceive it as not directly affecting them personally, and hence some lack a sense of urgency (\autoref{sec:beliefs}). Our participants then bear resemblance to users who are `engaged' by security~\cite{forget2016or}, who when alerted to there being a problem will seek out a solution.

\input{limitations}

\subsection{Recommendations}

From our analysis, we provide the following Recommendations:

\begin{itemize}

    \item \textbf{Adaptive notification channels.} An approach would be to find a manageable way to `ramp up' successive notifications to users at scale. However, any additional effort to encourage remediation across a sequence of notifications is borne by the ISP (who is already the stakeholder `taking charge' of the problem for lack of direct engagement by manufacturers). In many cases, our data shows that participants acted in effect immediately upon seeing `a' notification, albeit the last in a series of similar notifications. One approach might then be to consider other channels after the first notification (as seen in \cite{cetin2017make}). 
    
    \autoref{fig:survival} also indicates that there may be diminishing returns for solely email notification (the QSnatch curve flattens out as time goes on).  Our data also showed that many participants acted on or close to the first notification they received. Email was an appropriate channel through which to reach many, but some users may need an alternative notification rather than a repeat notification. Email as a notification channel works for some, but alternatives should be explored (within cost-effectiveness for an ISP), for instance, quarantining the connection, voice mail, direct phone call or letter to the customer. 
    
    \item \textbf{Framing and planning within remediation notifications.} All but one of our participants acted on a notification that they received. There were issues for several participants in terms of deciding to act on a notification and then having to find an opportunity to enact the instructions. The notifications we studied here act as a reminder to imply that immediate action is needed, primarily due to evidence of an active malware infection. 
    
    A balance may be struck between this and the use of commitment devices in reducing postponement (as explored elsewhere for security update behaviors \cite{frik2019promise}), for example having a user set a reminder for themselves for the same evening or the next day. 
    Framing is also important, where a few participants presumed the email notification was fake at first. This relates to messenger effects in effective communication of ideal security behaviors \cite{briggs2017behavior}.
    
    \item \textbf{Tailor-made advice.} Advice was specific to QNAP, and specific to QSnatch infection, rather than requiring a diagnostic analysis to determine which steps to selectively apply, as per the recommendation of the manufacturer. It was thereby actionable, from the users who did not have IT experience (n=15), 9 (60\%) reported following the advice (5 reported asking for help), and one user reported following QNAP steps. 
    
    Consistent with what \cite{redmiles2020comprehensive} recommended, we found evidence that minimum and concise instructions work, when measured via the remediation of IoT malware infection. Most of the time, Internet Service Providers (ISP) are restricted by laws and regulations, such as the General Data Protection Regulation (GDPR), from collecting data on the population of user devices in the local network. So in many cases, they cannot know in advance which is the infected device to provide tailor-made advice, thus we have to rely on generic advice for most cases. An intermediate point can be gaining consent from users to actually identify the infected device in their network to offer accurate help.

\end{itemize}

%% file: limitations.tex
\subsection{Limitations and future work}
\label{sec:limitations}

A limitation of our methodology is that study was carried out in a single Internet Service Provider (ISP) that has an established process for notifying users. Thus more research might be necessary to compare how this process happens in different ISPs. Additionally, we focus on a single persistent IoT malware family as a case study, QSnatch. Other persistent malware families might require different steps, and they might be harder to remove \cite{bock2019assessing, brierley2020persistence}. Thus, future research could consider comparing different persistent IoT malware families, to understand the applicability of our findings to other cases.

As with previous work \cite{rodriguez2021user, bouwmeester2021thing}, the results of this research were based on users self-reported behavior during interviews. The interviews were performed after a certain period of time. Ideally, we would have contacted users close to the notification time. However, the ISP was already notifying customers as part of their abuse handling process as explained in the \autoref{sec:methodology}. One of the authors was embedded in the ISP for a period of time; thus, we used historical data that allowed us to observe the QSnatch malware behavior over the period of a year. By using a larger historical timeframe, we could include a larger sample of affected users at the cost of more time between the remediation and the interview. If we wanted to time the interview close to the notification, then we would have to accept a much smaller user sample. In our results we found that 92\% of the participants stated that they received the notification, and all participants recalled what they did with it; thus, there is no evidence that they forgot what they did. This is in line with earlier work. Studies of security experiences, such as software updates \cite{vaniea2016tales} and social diffusion of security-related behaviors \cite{das2014effect,das2019typology}, gathered insights on user behaviors across potentially far-reaching timescales.

Finally, due to the manual intense process of retrieving the notification(s) dates, we could not compare the cleanup time of QSnatch with Mirai and QSnatch, thus future research could look into that. We only interviewed QSnatch infected users, thus we learned about their process of handling the infection. Previous work \cite{Cetin2019, rodriguez2021user, bouwmeester2021thing} has looked into the remediation process of Mirai, thus we focus on a group that has never been studied before, victims of persistent malware.

%% file: conclusion.tex
\section{Conclusion} \label{sec:conclusion}

Internet Services Providers use different methods to communicate with infected subscribers, and according to best practices~\cite{rfc6561}, email notifications is one of them. This paper shows that notifications play a crucial role in the cleanup process of persistent malware like QSnatch. In contrast to previous predominant malware families (e.g., Mirai or PC malware), an automatic scan or an AV tool or power cycle do not get rid of the QSnatch malware. As we observed in \autoref{fig:survival} QSnatch takes a longer time to get clean. 

Does QSnatch take longer to clean because it is hard? The remediation advice of the ISP consisted of a convoluted series of steps, however, most users reported having high technical competency.  Participants did not find following the steps as a problem, there might be a self-selection process of users with some IT experience, so these users might be comparing the steps to IT tasks. Hence, we dealt with the idealized user, they are capable, they are motivated, but it clearly takes time to organize cleaning of the device(s), and the majority of users had to receive multiple notifications to prompt them to act. The necessity of an external prompt for them to act contributes the non-trivial nature of QSnatch remediation.

In this study, we found that there is a lack of feedback loop about infections and cleanup success. Users had to be notified in order to act. An external party, in this case, the ISP, had to tell users they are infected and provide tailor-made advice to execute the right steps. This is not always possible for the ISPs since they cannot know in advance which malware and which device has been infected. 

Nevertheless, our study shows that all users remediated at some point, so damage is less with this self-selected user. It can happen that this will fall apart when average users use products affected by persistent malware, but it can also be that manufacturers such as QNAP are building already tools similar to Windows tools such as malware scanners and automatic updates from which average users will benefit.

In this study we have also found out, similar to~\cite{ cetin2019tell}, that email notifications could be effective. During circumstances such as a global pandemic where users depend on their Internet connection, this can be a good alternative. Similar to \cite{rodriguez2021user, bouwmeester2021thing} we found that when users are informed of a security issue, they are willing to act. They take responsibility although they do need some time and effort to execute the steps. In this, however, we need to take into account that technical abilities are key, to comprehending the notification, understanding what needs to be done, and knowing how to do it in a sufficiently complete and error-free manner. Connecting this thread of interdependent activities was not difficult for our participants, but it may be for those who are less tech-savvy. This is especially important in the absence of direct indicators from smart devices as to their security status, as explained in the opening arguments of this paper.

\label{sec:conc}

Regarding future work, we found that participants who took a long time to remediate their devices had generally received the highest number of successive notifications from the ISP. More correlation of technical and qualitative data is required to understand the role of communications and communication channels, and users planning strategies, especially as persistent malware continues to affect consumer devices.

%% file: acknowledgements.tex
\section*{Acknowledgments}\label{sec:ack}
\small
The authors would like to thank our anonymous reviewers for their feedback and suggestions to improve the quality of our manuscript. We thank the partner ISP company for permitting access to the abuse department data and knowledgeable staff, and facilitating engagement with their customers. This publication is part of the MINIONS project (grant no. 628.001.033) of the ``Joint U.S.-Netherlands Cyber Security Research Programme'' which is (partly) financed by the Dutch Research Council (NWO); and the ``Hestia Research Programme'' (grant no. VidW.1154.19.011).

%% file: appendix.tex
% % !TeX root = paper.tex
\clearpage
\onecolumn
\section{Appendices}

\subsection{Email notification content}\label{sec:appendixA}

\noindent\fbox{%
    \parbox{\textwidth}{%
What's going on and how can I fix it?

A NAS from the supplier QNAP connected to your Internet connection is infected with the QSnatch malware. This infection poses a major risk to the safety of your files on the device. It is important to manually update your NAS operating system and malware remover app. Use the steps below: \vspace{0.5cm} 

Operating system:
\begin{itemize}
\item Go to the website: \url{qnap.com/en-en/download}
    \item Under ``1 - Product type", select the option ``NAS / Expansion" and select the number of slots present on the right.
    \item Under ``3 - Model", select the type of NAS you are using.
    \item Under the ``Operating System" tab, select the most recent version and download it via the ``[REGION]" button.
    \item Open the NAS on your PC or Mac and choose firmware update, and then Manual update.
    \item Browse to the downloaded file and update the firmware / operating system.
\end{itemize}

Malware Remover app:
\begin{itemize}
    \item Go to APP Center and choose ``Malware Remover" and download it on your PC or Mac.
    \item Click on ``manual update" in App center, browse to the download file and update the Malware Remover.
    \item Run a scan with the Malware remover.
\end{itemize}

\vspace{0.5cm} 
What happens if I don't do anything?

The security problem on your Internet connection is a major threat. If you do not perform the steps or do not perform them correctly, we may place your Internet connection in our secure environment (quarantine). You can then temporarily make limited use of your Internet connection. By doing this we also protect your personal files and data.
\\\\ 
Do you have any questions? Then you can ask this in a reply to this e-mail. 
}}

\newpage
\subsection{Interview Questions} \label{sec:appendixB}
\begin{table}[h]
\scriptsize
\centering

\label{table:questionsinterview}
\begin{tabular}{ p{2cm} p{6cm}|p{6cm}}

&   \textbf{Did perform the steps}                                                 & \textbf{Did not/partially perform the steps}                                            \\
\midrule[2px]
Check questions  &  Are you the person who manage the QNAP device? & Are you the person who manage the QNAP device?
                
                \\
              
                & Do you use the device for business or private purposes?  & Do you use the device for business or private purposes? 
                                                           \\\\
  \midrule                                                                                
 Opportunity &Did you receive the notification email?                                         & Did you receive the notification email? \\\\
 
 &Did you do the steps in notification email?                                         &  Did you do the steps in notification email? (if not) what did you do? \\\\

   &                          Did you use any tools to perform the steps?                                     & Did you lacked any tools to perform all of the steps?                                                  \\\\
  &                          Did you have enough time to perform the steps?                                  & Did you not have enough time to perform the steps?                                                   \\\\
   &                         Was the location of the device or any of the tools you used an issue to access it?                                  & Was the location of the device or any of the tools you used an issue to access it?                                                  \\\\
                                                             
 & Have any people helped you perform the steps?                                & Did any people try to help you perform the steps?                                                 \\\\
    &                        Do some people you know have a strong opinion on performing the steps?          & Do some people you know have a strong opinion on performing the steps?
                            \\\\

\midrule

Capability & Did you understand the steps?                                                    & Did you understand the steps?  \\\\                                                                         
 &Did you find the steps challenging? &

Did you find the steps challenging? \\\\

  &  Did you have any physical or bodily limitations that made the steps challenging? & Did you have any physical or bodily limitations that prevented you from finishing the steps? \\\\

 & Can you give a rough indication of
how much time it took to complete the steps? &  Can you give a rough indication of how much time it took to complete what you did?  \\\\

                           & Did you know what malware is?      & Did you know what malware is?         \\\\
                         &  Did you know the difference between persistent and non-persistent malware?      & Did you know the difference between persistent and non-persistent malware?       \\\\
                          &  Did you think you could perform the steps?                                      & Did you think you could perform the steps?                                      \\\\
                           & Did you have previous experience with IT systems?                                      & Did you have previous experience with IT systems?                                                                      \\\\
& Did you find the steps useful?                                                   & Did you find the steps useful?    
     
\\\\

\midrule
  Motivation &                          What do you think would happen if someone does not follow the steps?                                 & What do you think would happen if someone does not follow the steps?                                                                \\\\
   &                         Did you think you are responsible for performing the steps?                     & Did you think you are responsible for performing the steps?                                     \\\\
                                                                                                                    
 &What did you feel while you performed the steps?                              & What did you feel when you received the notification email?                                   \\\\
 
  &                          Did an impulse helped you perform the steps?                                  & Did an impulse prevent you from performing the steps?                                         \\\\
                                                                                         
 \midrule
 Exit question    &                       Is there anything that you would like to add that is relevant and we did not ask?  & Is there anything that you would like to add that is relevant and we did not ask?\\
\bottomrule[1px]

\end{tabular}

\end{table}

\scriptsize
Note: Before the interview started, the researcher carrying out the interview took time to introduce himself, provided a description of the research, and asked for consent to proceed with the interview and data collection. Before the exit question, some demographic questions were asked, specifically self-reported gender and age.
\\